\documentclass[10pt, conference, letterpaper]{IEEEtran}

\usepackage{color}
\usepackage[american]{babel}

\usepackage{cite,comment,subfigure}

\ifCLASSINFOpdf
  \usepackage[pdftex]{graphicx}
  \graphicspath{{./}}
  \DeclareGraphicsExtensions{.pdf,.eps}
\else
  \usepackage[dvips]{graphicx}
  \graphicspath{{./}}
  \DeclareGraphicsExtensions{.eps, .png}
\fi

\usepackage{array}
\usepackage[cmex10]{amsmath}
\usepackage[hyphens]{url}
\usepackage{paralist}
\usepackage{amsthm}
\usepackage{amssymb}
\usepackage{pgfplots}
\usepackage{tikz}
\usetikzlibrary{shapes}
\usetikzlibrary{plotmarks}
\usepackage[ruled]{algorithm2e}
\usepackage[caption=false]{subfig}
\usepackage{epstopdf}
\usepackage{amsmath}

\newcommand\mynote[1]{} 

\newtheorem{proposition}{Proposition}

\begin{document}


\title{Cost-aware caching: optimizing cache provisioning and object placement in ICN}

\author{
  \IEEEauthorblockN  {
		Andrea Araldo\IEEEauthorrefmark{1}\IEEEauthorrefmark{2}, 
		Michele Mangili\IEEEauthorrefmark{1}\IEEEauthorrefmark{3}, 
		Fabio Martignon\IEEEauthorrefmark{1}\IEEEauthorrefmark{4} and 
		Dario Rossi\IEEEauthorrefmark{2}  }
\begin{tabular}{cc}
  \IEEEauthorrefmark{1} LRI, Universit\'e Paris-Sud, \{first.last\}@lri.fr &      \IEEEauthorrefmark{3} DEIB, Politecnico di Milano \\    
  \IEEEauthorrefmark{2} Telecom ParisTech, \{first.last\}@enst.fr &
  \IEEEauthorrefmark{4} IUF, Institut Universitaire de France\\
\end{tabular}
} 

\maketitle

\begin{abstract}
Caching is frequently used by Internet Service Providers as a viable technique to reduce the latency perceived by end users, while jointly offloading network traffic. While the cache hit-ratio is generally considered in the literature as the dominant performance metric for such type of systems, in this paper we argue that a critical missing piece has so far been neglected.
Adopting a radically different perspective, in this paper we explicitly account for the \emph{cost} of content retrieval, i.e. the cost associated to the external bandwidth needed by an ISP to retrieve the contents requested by its customers.  Interestingly, we discover that classical cache provisioning techniques that maximize cache efficiency (i.e., the hit-ratio), lead to suboptimal solutions with higher overall cost.
To show this mismatch, we propose two optimization models that either minimize the overall costs or maximize the hit-ratio, jointly providing cache sizing, object placement and path selection. We formulate a polynomial-time greedy algorithm to solve the two problems and analytically prove its optimality. We provide numerical results and show that significant cost savings are attainable via a \emph{cost-aware}   design.
\end{abstract}

\section{Introduction}
It has been extensively observed that contents in the Internet are characterized by a certain popularity skew: a relatively small number is requested very frequently while, on the other hand, a long tail of objects are rarely accessed~\cite{Breslau1999}. As caching in Internet leads to significant benefits~\cite{Agyapong2012}, caches are not only currently widely deployed (e.g., Web proxies, Content Delivery Networks CDNs) but also expected to play a paramount role in future Internet architectures (e.g., Information Centric Networks ICNs). 

Apart from the specific caching solution, it is widely accepted that caching is profitable for Content Providers (CPs), Internet Service Providers (ISP) and end users~\cite{Agyapong2012}. It makes CPs reduce their servers' load, ISPs offload their networks and end users experience lower latency by accessing a closer copy of the requested contents. While caching problems have been explored by a massive amount of scientific literature, at the same time the economic implications of caching are usually neglected by current work~\cite{Tyson2012,Rossi2012,Chai2012,Rossini2013,Wang2013,Baev2008,Tuncer2013} and have so far received only marginal consideration~\cite{Baev2008,Applegate2010,Mangili2013}. Furthermore, even work addressing economic aspects limitedly focuses on the interaction between independently administered caching systems, thus leaving out the cache design issues that arise inside each system. 

Therefore, in the literature the cache system design and the economic implications of caching have been treated as orthogonal subjects: in this work we show instead that these two aspects are tightly coupled,  and should be jointly considered. For this reason, we do not consider the \emph{cache hit ratio} as our sole performance indicator, but also include the  \emph{cost of content retrieval}  that ISPs incur when, to serve their  customer requests, they have to retrieve objects from other ISPs, paying for the generated traffic.

As it is particularly appealing from an ISP point of view~\cite{Wang2013,Dan,Cho2011,Agyapong2012}, we study a caching system as a means to reduce this cost, and analyze how this objective impacts the caching system design.  To the best of our knowledge (Sec.~\ref{sec:related_work}), we are the first to relate the economical benefits of caching with the technical design decisions of the caching system inside the network. 
Summarizing our contributions:
\begin{enumerate}[(1)]
	\item We provide two optimization models that jointly consider cache sizing, object placement and path selection to optimize, respectively, the cost of retrieval vs. the hit-ratio. We define these models in terms of constraints and objectives respectively in Sec.\ref{sec:scenario} and Sec.\ref{sec:optimization_goals}.
	\item We analytically find the optimal system design for the above problems, (namely the optimal cache size, object placement and path selection). We do so via a greedy algorithm, of which we prove the optimality, which is also suitable for large scale scenarios (Sec.~\ref{sec:greedy_algorithm}).
	\item We show that the classic performance indicators are not informative enough of the cost of content retrieval and that focusing on the optimization of those indicators is detrimental with respect to costs. We therefore quantify and discuss the gains that ISPs would attain via  optimal cache design under different settings (Sec.~\ref{sec:numerical_results}).
\end{enumerate}

\noindent Although the above findings are general, we primarily envisage their application to an ICN network directly administered by the ISP, since the optimal cache design that we discover exploits the ubiquitous caching of ICN by distributing the cache closer to the most expensive links for the ISP.

%
%


\section{Related Work}
\label{sec:related_work}

\subsection{Economic implications of caching}

As caching literature is abundant, we first focus on the significantly smaller fraction that, as our work, considers economic aspects of caching. From a high level viewpoint, some works \cite{Dan,Cho2011,Agyapong2012,Wang2013,Tyson2012} 
qualitatively observe that caching can potentially reduce ISP costs by limiting inter-ISP traffic. Along this line, this work complements such qualitative observations by analytically quantifying the benefits of caching in terms of the overall cost savings.


A game theoretic perspective on the economic implications of caching is addressed in \cite{Chun2004,Kocac2013,Pham2013,Lee2012}. In more details,
\cite{Chun2004,Kocac2013,Pham2013} model a network of independently administered caching systems whose aim is to minimize costs, through a game theoretical approach and study the Nash equilibria. The definition of cost given in \cite{Chun2004} is, as admitted by authors, rather abstract and cannot represent a real monetary cost incurred by an ISP -- which, on the contrary, is the focus of our work. Similarly, results in \cite{Kocac2013,Pham2013} are valid under newly proposed pricing schemes, while our interest lays in studying the economic implications of caching under the current unchanged pricing model. 
Finally, \cite{Lee2012} shows that an ISP deploying its own caching system, thus being able to provide a better service, can get a larger market share and increase the revenues -- but does not otherwise consider the savings in terms of retrieval cost.

%

With the exception of \cite{Cho2011,Wang2013},  all the above work considers ISPs as atomic entities, focusing only in the interactions among them: hence, they neglect the design of the cache network, which is instead among our goals. Moreover, whilst \cite{Cho2011,Wang2013} study the internal network design, they however limitedly consider the reduction of inter-domain traffic, measuring it blindly across all the external links,  while we show that it is crucial to consider the relative differences of prices among each of these links. 
It follows that neither \cite{Cho2011} nor \cite{Wang2013} propose intra-ISP strategies specifically designed to decrease the content retrieval cost, that, to the best of our knowledge, we are the first to propose.

\subsection{Design of caching networks}

\noindent From a broader perspective, a vast literature has already delved into many aspects of caching:
\begin{enumerate}[(1)]
	\item \emph{cache sizing}\cite{Mangili2013,Rossi2012,Tyson2012}, i.e. where and how much cache should be installed;
	\item \emph{object placement}\cite{Mangili2013,Wang2013,Baev2008,Tuncer2013,Applegate2010}, i.e. which object to store and where; 
	\item \emph{path selection}\cite{Mangili2013,Rossini2013}, i.e. which sequence of nodes a request for a content should pass through;
	\item \emph{replacement policy}\cite{Podlipnig2003}, i.e. which object to evict in a cache if a new one has to be stored;
	\item \emph{decision policy}\cite{Laoutaris2004,Rossini2013,Chai2012}, i.e. choosing whether a new object should be cached or not.
\end{enumerate}

All the above work employs cost-unaware performance indicators, namely the overall network \emph{hit-ratio} \cite{Tyson2012,Rossi2012,Chai2012,Rossini2013}, the number of \emph{hops} \cite{Tyson2012,Wang2013,Chai2012,Baev2008,Rossini2013} or the \emph{link load} \cite{Tuncer2013,Mangili2013}. Conversely, this work explicitly contrasts cost-unaware vs. cost-aware cache designs.

As far as the methodology is concerned, optimization models \cite{Baev2008,Applegate2010,Mangili2013} have been proposed for caching but none of them tackles the cost of content retrieval. In addition, since ILP models are computationally expensive, \cite{Baev2008} and \cite{Applegate2010} propose approximation methods 
while \cite{Mangili2013} obtain numerical results under simplifying assumptions on the object catalog. On the contrary, the greedy algorithm we propose here allows us to find an exact optimal solution using realistic object catalogs.



\section{System Model}
\label{sec:system_model}
In this section we first assess the nature of the economic interactions that ISPs have with other Internet players (Sec.~\ref{sec:interaction}). We then provide two optimization models subject to the same constraints (Sec.~\ref{sec:scenario}) but having the conflicting objectives of cost vs. hit-ratio (Sec.~\ref{sec:optimization_goals}). We finally provide a polynomial time greedy algorithm to solve these problems, and prove its optimality (Sec.~\ref{sec:greedy_algorithm}).

\subsection{Economic interactions}
\label{sec:interaction}
\begin{figure}[t]
    \centering
    \includegraphics[width=2.95in]{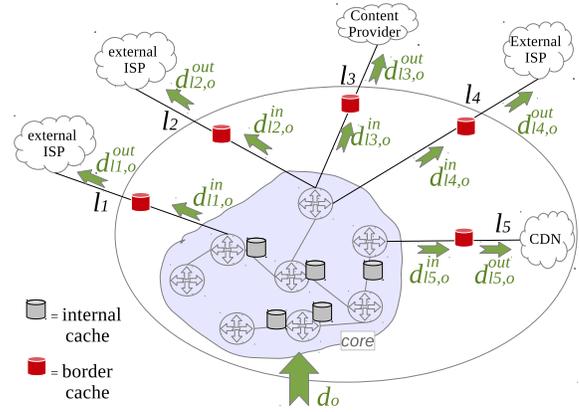}
    \caption{Model of the ISP. The ISP is connected to third party networks  through external links $l_1,\dots,l_5$. The arrows represent the flow of the object requests. Two sets of caches can be deployed in the network: border caches are installed in front of external links, whereas internal caches are co-located with routers. We denote with \emph{core} the set of all internal caches.}
    \label{fig:model}
\end{figure}
As shown in Fig.\ref{fig:model}, an ISP receives an incoming demand $d_o$, i.e. a sequence of requests for object $o$. The ISP serves them, retrieving $o$ through the available \emph{external links}, paying a traffic-related cost.  Internal links have a zero-cost, since the ISP does not have to pay to use them~\cite{Shakkottai2006}. Some of the requests can be served by caches in the ISP network. Therefore, the incoming demand $d_o$ is ``filtered'' by the caches and the demand on the external links (represented by $d_{l_1,o}^{out},\dots,d_{l_5,o}^{out}$) is less than $d_o$. The primary goal of an ISP is to minimize the cost associated to the external links utilization, by installing a limited amount of cache storage within its network.

As an example, consider a local ISP, that generally receives requests from end users and retrieves objects from an upstream ISP. Nonetheless, as thoroughly explained in \cite{Shakkottai2006,Dhamdhere2010}, there might exist more heterogeneous interactions. For instance, requests might also be served by peer local ISPs, CDNs or Content Providers (CPs) directly connected to the ISP, as in case of multihoming~\cite{Hau2011a}. We abstract the different types of interactions by distinguishing three categories of links based on the cost associated to the traffic flow:
\begin{enumerate}[(1)]
	\item \emph{Settlement-free peering link}: the ISP does not pay to use this link; usually, these links connect ISPs of the same tier.
	\item \emph{Provider link}: the ISP pays to receive traffic from it; example is a transit link to a higher-tier ISP. 
	\item \emph{Customer link}: the ISP is paid
\footnote{Usually, CDNs pay ISPs only in case ISPs are sufficiently large. In the other cases, settlement-free agreements are established \cite{Kramer2013,Agyapong2012}}
to receive traffic through this link; examples are the links toward   lower tier ISP, or CPs in multihoming~\cite{Hau2011a,Lee2012} or CDNs nodes.

\end{enumerate}
\mynote{XXXX AA: Alcune volte e' necessario introdurre una metodologia nuova quando il proprio modello non matcha con niente di esistente, ad esempio perche' si prova ad essere piu' generali di quanto si trova in letteratura. Ad es in Dhamdhere2010 pag2.col2 loro introducono una loro terminologia XXXX}
%
%
%
%
\subsection{Optimization constraints}
\label{sec:scenario}
\begin{table}
  \caption{Summary of the notation used in this paper.}
  \label{tbl:notation}
  \centering
  \begin{tabular}{| c | l |}
    \hline
    \multicolumn{2}{| c |}{\textbf{Input parameters} } \\
    \hline
    $\mathcal{O}$ & Set of objects \\
    \hline
    $\mathcal{N}$ & Set of routers of the ISP network \\
    \hline
    $\mathcal{L}$ & Set of external links \\
    \hline
    $\mathcal{L}_o$ & Set of external links that give access to object $o$ \\
    \hline
    $\mathcal{Q}$ & Set of requests received by the ISP \\
    \hline
    $\mathcal{Q}_o\subseteq\mathcal{Q}$ & Set of requests for object $o$ \\
    \hline
    $C_{tot}$ & Total amount of cache in the ISP network \\
    \hline
    $d_o$ & Number of requests for object $o$ that the ISP has to satisfy\\
    \hline
    $p_l$ & The cost that the ISP has to pay to retrieve a single \\
		  & object through $l$\\
    \hline
    \multicolumn{2}{c}{} \\


    \hline
    \multicolumn{2}{| c |}{\textbf{Decision variables}} \\
    \hline
    $P$ & Function that associates each request to the path, \\
				  & i.e. the sequence of ISP routers to be traversed\\
    \hline
    $\rho_o$ & Fraction of requests for object $o$ that are actually satisfied \\
	{}		 & in the core\\
    \hline
    $d_{core,o}^{out}$ & Number of requests for object $o$ that are not \\
	{}				   & satisfied by any internal cache \\
    \hline
    $d_{l,o}^{in}$ & Number of requests for object $o$ that are not satisfied by \\
	{}			   &	any internal cache and are directed toward link $l$\\
    \hline
   $d_{l,o}^{out}$ & Number of requests for object $o$ that are not satisfied by \\
	{}			   &	any internal cache, are directed toward link $l$, and are not\\
	{}			   &	satisfied by the border cache on $l$, thus exiting the ISP through $l$\\
    \hline
    $c(q)$ & Cost that ISP incurs to satisfy the request $q$ \\
    \hline
    $c$ & Total cost of retrieval \\
    \hline
    $h$ & Overall hit-ratio \\
    \hline
    $\mathcal{C}$ & Set of objects that are stored in at least one of the \\
	{}	&	caches of the ISP \\
    \hline
    $cs_n$ & Size of the internal cache co-located with the router $n$\\
    \hline
    $cs_{l}$ & Size of the border cache placed in front of link $l\in\mathcal{L}$\\
    \hline
    $x_{l,o}$ & It is 1 if border cache on link $l$ is caching object $o$. 0 otherwise \\
    \hline
    $x_{n,o}$ & It is 1 if the internal cache co-located with router $n$ is caching \\
			  &		object $o$. 0 otherwise \\
    \hline
    $l_o$ &  The cheapest among the links $\mathcal{L}_o$ that give access to object $o$\\
    \hline
    $l(q)$ & External link to which the request $q$ is directed\\
    \hline
    $pc_o$ & Potential cost of an object: $pc_o= d_o \cdot p_{l_o}$\\
    \hline
  \end{tabular}
\end{table}
Tab.~\ref{tbl:notation} summarizes the notation used in this paper. 
Let $\mathcal{O}$ be the object catalog and $\mathcal{L}$ the set of external links. Since our study pertains to costs, we do not consider the customer links, that represent income rather than costs. Indeed, if an object is accessible through one of these links, as observed by \cite{Agyapong2012,Kocac2013,Tyson2012}, the ISP has no interest in caching it since this would imply a loss of income. 
%
Without loss of generality, we therefore ignore customer links and consider $\mathcal{L}$ as the set of provider links and settlement-free peering links, and $\mathcal{O}$ as the set of objects that can be retrieved only through these links.
As depicted in Fig.~\ref{fig:model}, the ISP can install a cache in front of an external link $l$ (we call it \emph{border cache} on $l$ for brevity), but  can also place caches co-located with some router $n\in\mathcal{N}$ (we call this cache an \emph{internal cache}, and further denote with \emph{core} the set of all internal caches).
We next consider two sets of binary variables for object placement: $x_{l,o}$ and $x_{n,o}$. We set $x_{l,o} = 1$ if object $o \in \mathcal{O}$ is cached at the border cache on external link $l$. Similarly, $x_{n,o} = 1$ if $o$ is cached at the internal cache co-located with router $n$.

Let $\mathcal{Q}$ be the set of requests that the ISP receives. A path selection function $P$ associates to each request $q\in\mathcal{Q}$ a sequence $P (q)=\left[n_1,n_2,...,n_k,l(q)\right]$, where $n_1$ is the ingress router, i.e. the first ISP router traversed by $q$ and $l(q)$ is the external link to which $q$ is directed. Suppose that $o$ is the object requested in $q$. In case there exists a router $n$ inside the path $P(q)$ whose internal cache is storing $o$, i.e. $x_{n,o}=1$, then the request will be satisfied by the core. Otherwise, the same request will be forwarded to one of the external links $l$. Denoting with $d_o$ the amount of requests for object $o$ that the ISP receives, we are interested now in the fraction $\rho_o$ of them that are satisfied by the core. Knowing i) the set of incoming requests $\mathcal{Q}$, ii) the path selection function $P(\cdot)$ and iii) the object placement in the internal caches (expressed by the variables $x_{n,o}$), we can exactly compute the fraction of incoming demand for $o$ that is satisfied by the core as a function of these three data:
\begin{equation}
	\label{eq:rho}
	\rho_o = F  \left( 
					\mathcal{Q}, P,  \left\{ x_{n,o} | n\in\mathcal{N} \right\}
				\right)
\end{equation}
We anticipate that, since optimal solutions have no internal caches, we have that $\rho_o=0$ irrespectively of $F(\cdot)$, whose precise definition can be thus disregarded in what follows.
Let us denote with $d_{core,o}^{out}$ the number of requests that are not satisfied by the core:
\begin{equation} 
	d_{core,o}^{out} = (1-\rho_o) d_o \label{eq:core_as_filter} 
\end{equation}
\noindent where, in other words, the core acts as a filter for the incoming demand. Then, the demand $d_{core,o}^{out}$, that has not been served by the core, is spread over the external links. We denote with $d_{l,o}^{in}$ the part of this demand that is directed to link $l$. 
\begin{equation} d_{core,o}^{out} = \sum\limits_{l\in\mathcal{L} } d_{l,o}^{in} \end{equation}
Note that an object $o$ can be accessed only through a subset $\mathcal{L}_o$ of external links:\footnote{We also exclude from $\mathcal{L}_o$ the links that is not possible to use to retrieve $o$ because they are forbidden by some BGP policy.}
 It is impossible to retrieve $o$ through all the other links, hence:
\begin{equation} d_{l,o}^{in} = 0, \forall l\in\mathcal{L}\setminus\mathcal{L}_o \end{equation}
If $x_{l,o}=1$, the border cache on link $l$ is caching object $o$ and directly serves the demand $d_{l,o}^{in}$ that it receives. Otherwise the demand will exit the ISP. Therefore, denoting with $d_{l,o}^{out}$ the demand for $o$ flowing out through link $l$, we have:
\begin{equation} d_{l,o}^{out} = (1-x_{l,o}) \cdot d_{l,o}^{in} \end{equation}
Considering an external link $l$, we denote with $cs_l$ the size of the border cache installed in front of that link. Similarly, considering a router $n$, $cs_n$ is the size of the internal cache co-located with $n$. Cache sizes indicate the number of objects that can be stored.
It is straightforward to observe that:
\begin{flalign} 
	\sum\limits_{o\in\mathcal{O} } x_{l,o} = cs_l,  & \forall l\in\mathcal{L}
	\label{eq:border_cache_sizing} \\
	\sum\limits_{o\in\mathcal{O} } x_{n,o} = cs_n,  & \forall n\in\mathcal{N}
\end{flalign}
Finally, we suppose the total amount of caching storage that can be installed in the network to be upper-bounded  by $C_{tot}$:
\begin{equation} 
	\sum\limits_{l\in\mathcal{L} } cs_l + \sum\limits_{n\in\mathcal{N} }cs_{n,o} \leq
	C_{tot} .
	\label{eq:cache_budget}
\end{equation}
We assume that the bandwidth consumed to transmit a request is negligible compared to the requested object itself. Therefore, the costs that we consider account only for data traffic, while we assume to be zero the cost of request transmission, as in \cite{Pham2013}.
In addition, we also assume the object size to be the same for all objects, following the convention in the literature (as in \cite{Rossini2013,Chai2012,Mangili2013}), so that the bandwidth consumed for their transmission is always the same.
With these assumptions, we can define $p_l$, the price of link $l$, as the traffic cost that the ISP has to pay to retrieve a single object through it. 
Note that $p_l>0$ in case of provider link and $p_l=0$ in case of settlement-free peering link.
%
%
%
%
\subsection{Optimization goals}
\label{sec:optimization_goals}
As we previously argued, caching system performance may be expressed in terms of two equally relevant indicators, i.e. the \emph{cost of retrieval} \eqref{eq:cost_of_retrieval} and the \emph{hit-ratio} \eqref{eq:hitratio}:
\begin{equation} 
	\label{eq:cost_of_retrieval}
	c = \sum\limits_{\substack{o \in \mathcal{O} \\ l \in \mathcal{L}}}
			{p_l \cdot d_{l,o}^{out} } 
\end{equation}
\begin{equation}
	\label{eq:hitratio}
	h = 1-
		\sum\limits_{ \substack{ l\in\mathcal{L} \\ o\in\mathcal{O} } } 
				d_{l,o}^{out}  /
	  \sum\limits_{o\in\mathcal{O} } d_o
\end{equation}
An ISP may design the cache system in order to minimize the cost $c$ or to maximize the hit-ratio $h$. We model these two conflicting goals with two different multi-objective optimization problems, 
that we call respectively MIN-COST and MAX-HIT, both subject to constraints \eqref{eq:rho}-\eqref{eq:cache_budget}, and having, respectively, the following objectives:
\begin{equation}
	\label{eq:obj_cost_hitratio}
	\mbox{MIN-COST: } \min \left[ c, 1-h  \right]
\end{equation}
\begin{equation}
	\label{eq:obj_hitratio_cost}
	\mbox{MAX-HIT: } \min \left[ 1-h, c  \right]
\end{equation}
Note that the \emph{order} of the functions we minimize in the two problems is important: in MIN-COST the primary goal is the minimization of the cost while the secondary goal is the maximization of the hit-ratio (that we express as the minimization of the miss-ratio, $1-h$). On the contrary, in MAX-HIT the sequence of objective functions is inverted.


In both MIN-COST and MAX-HIT, the ISP must jointly solve the following design tasks:
\begin{enumerate}
	\item \label{itm:cache_sizing} \emph{Cache sizing}: allocating the total cache budget $C_{tot}$ among internal and border caches, fixing $cs_l$ and $cs_n$.
	\item \label{itm:path_selection} \emph{Path selection} represented by $P$: deciding how to route the requests inside the network.
	\item \label{itm:obj_placement} \emph{Object placement}: deciding which objects to store in each cache, fixing $x_{l,o}$ and $x_{n,o}$.
\end{enumerate}
\noindent The solution of the optimization problem yields to a network configuration that precisely corresponds to the above tasks.

%
%
%
%
%
\subsection{Greedy algorithm}
\label{sec:greedy_algorithm}
\begin{algorithm}[t]
	\small
	\caption{Greedy algorithm for MIN-COST}
	\label{alg:greedy_algo}
	  \SetKwInOut{Input}{Input}\SetKwInOut{Output}{Output}
	  \Input{ $\mathcal{N},\mathcal{L},\mathcal{O},C_{tot},p_l$ }
	  \Output{ $cs_l,cs_n,x_{l,o},x_{n,o}$ }
	  \nl $x_{l,o}\Leftarrow x_{n,o}\Leftarrow cs_{n}\Leftarrow\
			 cs_{l}\Leftarrow 0;$ 
			$\forall n\in\mathcal{N},\forall l\in\mathcal{L},\forall o\in\mathcal{O}$
			\label{lst:init}\\
	  \nl \ForEach{ $o\in\mathcal{O}$ } {
	  		\nl fix $l_o$ s.t. $p_{l_o}=\min\lbrace p_l|l\in\mathcal{L}_o \rbrace$;
				\label{lst:lo}\\
			\nl $d_{l_o,o}^{in}\Leftarrow d_o$; \,\,\,
				{ $d_{l,o}^{in}\Leftarrow 0$,\, 
				$\forall l\in\mathcal{L}\setminus\lbrace l_o \rbrace$}
				\label{lst:direct_do}\\
	  		\nl $pc_o \Leftarrow d_o \cdot p_{l_o}$ 
				//definition of \emph{potential cost}\
				\label{lst:potential cost}\\
		}
			\label{lst:impose_path}
	  \nl $\mathcal{O}_{sorted} = sort \left( \mathcal{O} \right)$
			\label{lst:sorting} \\
	  \nl $\mathcal{O}_{C_{tot}} \Leftarrow$ 
			the set of the first $C_{tot}$ objects of $\mathcal{O}_{sorted}$;
			\label{lst:obj_to_cache}\\

		\nl $x_{l_o,o} \Leftarrow 1,\,\forall o\in\mathcal{O}_{C_{tot}}$
			// Store $o$ in the cache on $l_o$
			\label{lst:impose_obj_placement}\\
		\nl $\sum\limits_{o\in\mathcal{O} } x_{l,o} = cs_l,\,\forall l\in\mathcal{L}$
			// Cache sizing
			\label{lst:impose_cache_sizing}
\end{algorithm}
We propose a greedy algorithm (described in Alg.~\ref{alg:greedy_algo}) which solves the optimization problem MIN-COST \eqref{eq:obj_cost_hitratio}. Step~\ref{lst:init} is the initialization. It is worth pointing out that at the end of the algorithm, $cs_n=0$ for all internal nodes $n$. This means that there are no internal caches and thus the incoming demand is not filtered by the core. Therefore, $\rho_0 = 0$ and thanks to \eqref{eq:core_as_filter}, $d_{core,o}^{out} = d_o$.
In step \ref{lst:lo} we associate to each object $o$ a link $l_o$ which is the cheapest among the links $\mathcal{L}_o$ that give access to $o$, i.e. the one with the minimum price (breaking ties arbitrarily). In step~\ref{lst:direct_do} we direct all the incoming demand only to the link $l_o$ chosen above.
In step~\ref{lst:potential cost} we define the \emph{potential cost} of each object $o$, that depends only on the demand $d_o$ and the price of link $l_o$ chosen above.
In step~\ref{lst:sorting} we sort objects in a descending order according to their potential cost. If there are objects with the same cost  $pc_o$, we sort them in a descending order based on their demand $d_o$. In step~\ref{lst:obj_to_cache}, we choose the objects that we want to cache, namely the first $C_{tot}$ objects of $\mathcal{O}_{sorted}$. In step~\ref{lst:impose_obj_placement} we place each of the selected objects in one border cache, namely the cache on the link $l_o$ chosen before. Finally, step~\ref{lst:impose_cache_sizing} simply corresponds to \eqref{eq:border_cache_sizing}.
Note that, referring to the design tasks outlined in Sec.~\ref{sec:optimization_goals}, step~\ref{lst:direct_do} solves the path selection task, step~\ref{lst:impose_cache_sizing} the cache sizing and step~\ref{lst:impose_obj_placement} the object placement.

\begin{figure*}[t]
	\subfigure[Relative cost saving of MIN-COST vs MAX-HIT. The (narrow) 95\% confidence band for each curve is plotted.]
		{\includegraphics[width=0.3\textwidth]{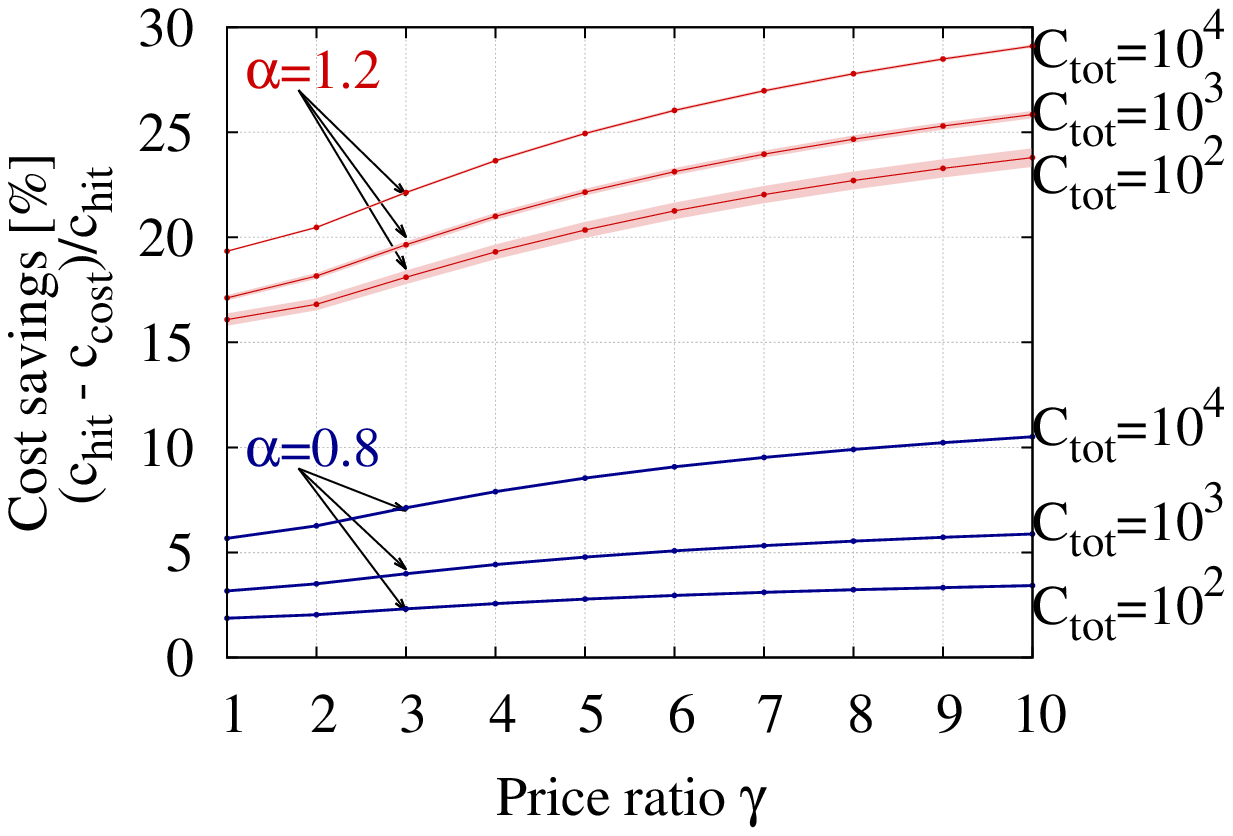}}
	\subfigure[Relative hit-ratio loss of MIN-COST vs MAX-HIT. For each $\alpha$, a band including the 95\% confidence interval of all curves is plotted.]
		{\includegraphics[width=0.3\textwidth]{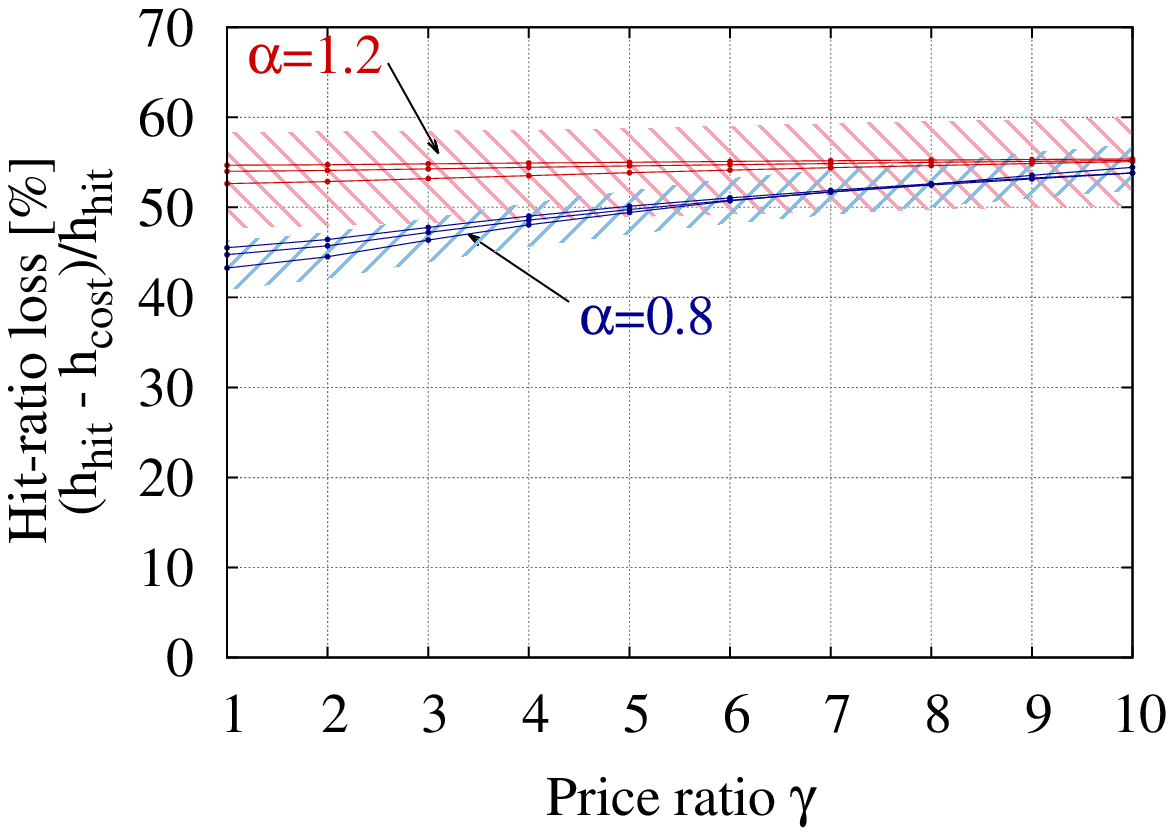}}
	\subfigure[Relative size of the two border caches for MIN-COST for $\alpha=1.2$. The 95\% confidence intervals are shown.]
		{\includegraphics[width=0.4\textwidth]{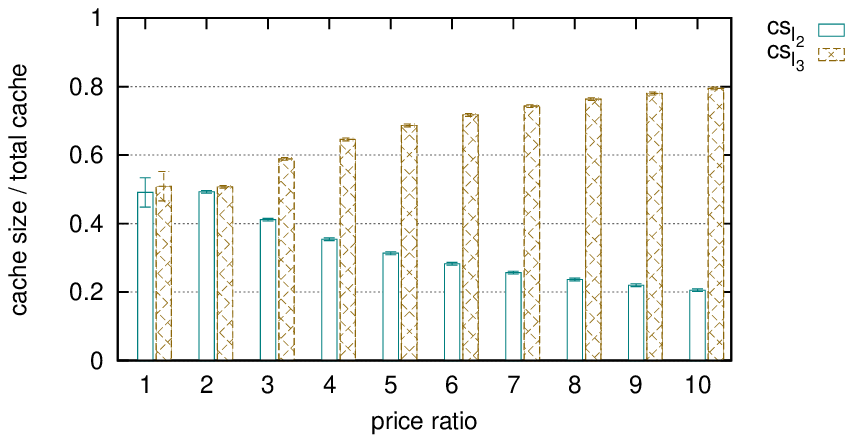}}
	\caption{Numerical results: (a) cost saving, (b) hit-ratio loss  and (c) cache size as a function of the price ratio $\gamma$}
	\label{fig:results}
\end{figure*}

\begin{proposition}
	\label{pro:greedy_cost}
	Alg.~\ref{alg:greedy_algo} gives the optimal solution of the optimization problem MIN-COST \eqref{eq:obj_cost_hitratio}.
\end{proposition}
\begin{IEEEproof}
To prove the proposition, we first find a lower bound $c_{LB}$ to the cost of retrieval and then show that the cost of the solution given by Alg.~\ref{alg:greedy_algo} is exactly $c_{LB}$. To find $c_{LB}$, we consider a generic network with a total cache budget $C_{tot}$.

In general, the path selection may make each request $q\in\mathcal{Q}$ for the same object $o$ follow a different path, for instance directing it through a different external link. Therefore, to satisfy each request $q$, the ISP may incur in a different cost $c(q)$. In particular, $c(q)=p_l$ if $q$ exits through external link $l$ while $c(q)=0$ if it is satisfied by any cache, either internal or border cache. The total cost of retrieval is the cost incurred to satisfy all the requests, and thus \eqref{eq:cost_of_retrieval} can be expressed also as:
	\begin{equation} 
		\label{eq:sum_of_sum}
		c=	\sum\limits_{q\in\mathcal{Q}} c(q) = 
			\sum\limits_{o\in\mathcal{O} }\sum\limits_{q\in\mathcal{Q}_o} c(q) 
	\end{equation}
where $\mathcal{Q}_o\subseteq\mathcal{Q}$ is the set of the requests for object $o$. Denote with $\mathcal{C}$ the set of objects that are stored in some internal or border cache. If  $o\notin\mathcal{C}$, each request $q\in\mathcal{Q}_o$ must exit the ISP through a link $l(q)\in\mathcal{L}_o$ with price $p_{l(q)}$: 
	\begin{multline}
		\label{eq:bound_stored} 
		\sum\limits_{q\in\mathcal{Q}_o} c(q)  = 
		\sum\limits_{q\in\mathcal{Q}_o} p_{l(q)} \geq
		\sum\limits_{q\in\mathcal{Q}_o} p_{l_o} =
		d_o \cdot p_{l_o} = pc_o\\
		\forall o \in\mathcal{O}\setminus\mathcal{C}
	\end{multline}
The inequality holds since, by definition, every possible link $l(q)$ has a price $p_{l(q)}$ that cannot be smaller than the price $p_{l_o}$ of the cheapest link $l_o\in\mathcal{L}_o$. The last equality follows from the definition of potential cost (step 5 of Alg.~\ref{alg:greedy_algo}). On the other hand, in case $o$ is stored in some caches, the path selection may be able to redirect some requests (but not necessarily all) to that copy, leading to a zero-cost. Therefore, irrespectively of path selection, the following holds:
	\begin{equation}
		\label{eq:bound_not_stored} 
		\sum\limits_{q\in\mathcal{Q}_o} c(q) \geq 0,
		\forall o \in\mathcal{C}
	\end{equation}
Given \eqref{eq:bound_stored} and \eqref{eq:bound_not_stored}, we can rewrite the \eqref{eq:sum_of_sum} as
	\begin{equation} 
		\label{eq:lower_bound_quasi} 
		c \geq\sum\limits_{o\in\mathcal{O}}     
			\begin{cases} 
			      0 & \mbox{if } o \in\mathcal{C}\\ 
			      pc_o & \mbox{if } o \in\mathcal{O}\setminus\mathcal{C}
		    \end{cases}=
		\sum\limits_{o\in\mathcal{O}\setminus \mathcal{C} } pc_o =
		\sum\limits_{o\in\mathcal{O} } pc_o - 
		\sum\limits_{o\in\mathcal{C}} pc_o 
	\end{equation}
The term $\sum\limits_{o\in\mathcal{C}} pc_o$ is the sum of the potential costs of a set of $C_{tot}$ objects. It
is upper bounded by the  potential costs of the most expensive objects, which are the ones in $\mathcal{O}_{C_{tot}}$ (as in step~\ref{lst:obj_to_cache} of Alg.~\ref{alg:greedy_algo}).
%
Hence, we can rewrite \eqref{eq:lower_bound_quasi} as follows:
	\begin{equation} 
		c \geq
		\sum\limits_{o\in\mathcal{O} } pc_o - 
		\sum\limits_{o\in\mathcal{O}_{C_{tot}}} pc_o =
		\sum\limits_{o\in\mathcal{O}\setminus \mathcal{C}_{tot} } pc_o =
		c_{LB}
	\end{equation}
The last equation shows that considering a generic network with total cache budget $C_{tot}$, irrespectively of path selection and objects placement, the cost of retrieval is no less than $\sum_{o\in\mathcal{O}\setminus \mathcal{C}_{tot} } pc_o$, which, therefore, is the lower bound $c_{LB}$ we were looking for. 
By simple calculation, that we are forced to omit due to lack of space, it is possible to show that the cost of retrieval provided by  Alg.~\ref{alg:greedy_algo} is exactly $c_{LB}$. 
For the sake of completeness, note that for objects having the same potential cost, thanks to the step~\ref{lst:sorting} of Alg.~\ref{alg:greedy_algo}, the most popular one is stored. This ensures that, among all the possible solutions at cost $c_{LB}$, Alg.~\ref{alg:greedy_algo} provides the one with the largest hit-ratio.
\end{IEEEproof}
%


Slightly modifying the previous algorithm, we can solve the optimization problem having the maximization of the hit-ratio as primary goal. The proof, which we omit for reason of space, follows similar arguments.
\begin{proposition}
	\label{pro:greedy_hitratio}
	An optimal solution of the MAX-HIT optimization problem \eqref{eq:obj_hitratio_cost} is given by Alg.~\ref{alg:greedy_algo}, modifying the $sort$ function (step~\ref{lst:sorting}), sorting objects in $\mathcal{O}$ in a descending order according to their demand $d_o$ and, if there are objects with the same demand, sorting them in a descending order based on their potential cost $pc_o$.
\end{proposition}


%
\section{Numerical results}
\label{sec:numerical_results}

We now provide and discuss numerical results, obtained through a Matlab implementation of our greedy algorithm, aimed at both (i) quantitatively assessing the economic savings that an ISP can get by using our proposed cost-aware cache design and (ii) understanding structural differences in terms of cache sizing between the cost-aware vs. cost-unaware designs.

While our framework is general and can be applied to every type of ISP, in this section we focus on a Local ISP with three external links: a peering link $l_1$, a ``cheap link'' $l_2$ and an ``expensive link'' $l_3$ with respective prices $p_{l_1}=0$, $p_{l_2}=1$ and $p_{l_3} = \gamma\cdot p_{l_2}$, where $\gamma\in\lbrace1,2,\dots,10\rbrace$ denotes the price ratio.  We consider a realistic Internet-scale catalog~\cite{Rossini2013}, consisting of $10^7$ objects whose popularity is Zipfian with exponent $\alpha=\lbrace 0.8, 1.2 \rbrace$\cite{Fayazbakhsh2013}. The total cache budget is $C_{tot}=\lbrace 10^2, 10^3, 10^4 \rbrace$ objects.
For each configuration we generate 40 scenarios. In each scenario, objects are assigned to each external link with 0.5 probability, so that an object can be reachable through more than one link. If there are unassigned objects, each of them is uniformly assigned to one of the links. We then compute the 95\% confidence intervals shown in the plots.
In each experiment, we calculate the cost-optimal configuration using the MIN-COST algorithm, obtaining an optimal total cost $c_{cost}$ (and a corresponding hit-ratio $h_{cost}$). We then calculate the hit-ratio-optimal configuration using the MAX-HIT algorithm, obtaining an optimal hit-ratio  $h_{hit}$ (and a different cost $c_{hit}$). We finally define the \emph{hit-ratio loss} as $(h_{hit} - h_{cost})/h_{hit}$ and the \emph{cost saving} as $(c_{hit} - c_{cost})/c_{hit}$ respectively.
The former expresses how much the hit-ratio degrades in the MIN-COST with respect to the MAX-HIT configuration. The latter, instead, gauges the cost savings of MIN-COST that are lost in the MAX-HIT solution.

%
%
%
%
%
%
\subsection{Cost savings}
\label{sec:cost_savings}
From Fig.~\ref{fig:results}-(a) and (b) we observe that, as the price ratio increases, optimizing the cost enables cost savings of up to 30\%.
As an expected side effect, this induces a loss of caching efficiency in terms of cost-blind metrics (i.e., the hit-ratio) up to 60\%.
Otherwise stated, if an ISP wants to get economical benefits from the use of caches, hit-ratio optimization must be a secondary objective: indeed, a loss of hit-ratio efficiency can translate into significant gains in terms of cost savings.  Moreover, this holds especially when the external links are very heterogeneous in terms of prices (which we expect to be the common case), since the cost savings increase as the price ratio increases.

Note that both cost savings and hit-ratio loss consistently increase with the popularity skew (from $\alpha=0.8$ to $1.2$).
In addition, the cost savings also increase consistently as the total cache size $C_{tot}$ increases from $C=10^2$ to $C=10^4$ (see the different curves in the plot). Intuitively, this means that when the cache budget is high, the operator has more freedom in its allocation, and the potential gains are larger; as a consequence, choosing the right cache planning strategy is crucial for attaining these potential gains. On the other hand, the hit-ratio loss is quite unaffected by $C_{tot}$ (Fig.~\ref{fig:results}-(b)).

\mynote{XXXX AA:Interessante capire cosa cambia al variare di alpha XXXX}
%

\subsection{Cache sizing}
\label{sec:cache_sizing}

We analyze how the optimal cache planning varies depending on the price ratio $\gamma$.
As expected, the MIN-COST cache sizing places no cache on the peering link $l_1$, while it proportionally allocates more cache on the more expensive links. Fig.~\ref{fig:results}-(c) shows the percentage of cache budget allocated to each of the two border caches behind the cheap link $l_2$ and the expensive link $l_3$ (normalized so that their sum equals 1). The shown cache sizing permits to store the objects that would require a higher cost of retrieval.
The overall results depend on a combination of orthogonal factors such as prices, content popularity and content availability behind each link.

At first, observe that each object may in theory be reachable through both links. Step~\ref{lst:impose_path} of Alg.~\ref{alg:greedy_algo} ensures that, in this case, the copy on the most expensive link is never used. This implies that the cheap link $l_2$ is more exploited since it is used to retrieve the objects that it is the only one to provide, as well as the objects that are provided by both links. On the contrary, $l_3$ is used only to retrieve objects that it is the only one to provide. As a consequence, more objects are retrieved through $l_2$ rather than $l_3$.
For this reason, for small values of price ratio, a non-negligible cache size is allocated on $l_2$, which is the most used, even if it is cheaper: when prices are homogeneous, cache sizing is impacted by content availability more than by prices. On the contrary, when prices' heterogeneity grows, its influence prevails: the cache allocated on $l_3$ noticeably stands out , even if $l_3$ is less used. In this case, the cache sizing is driven by prices more than by content availability.
As for the content popularity impact, while Fig.~\ref{fig:results}-(c) only shows the results for $\alpha=1.2$, we verify that for $\alpha=0.8$, the trend remains the same, but the difference between $cs_{l_2}$ and $cs_{l_3}$ is more evident: the less the skewness in the popularity, the more the cache sizing is impacted by the prices.
%

\section{Conclusion}
In this paper we analyzed caching as a means to limit the \textit{cost of content retrieval} incurred by ISPs. Particularly, to the best of our knowledge, this work is the first to consider how the economic implication of caching may impact the design of a caching system. 

Specifically, we formulated two optimization problems that either maximize the hit-ratio (MAX-HIT), as usually done in the literature, or directly minimize the cost of retrieval (MIN-COST). We are able to solve large-scale instances of these problems with a greedy algorithm, that we analytically prove to give an optimal solution. Contrasting numerical solutions of MIN-COST and MAX-HIT, we find that the additional cost in the classical approach (MAX-HIT) grows with both (i) the price diversity (ii) the cache budget. 

Our work shed new light on the caching problem, as it revisits it under a new viewpoint. Our results suggest that new approaches are required for the design of a cost-aware distributed caching system. Indeed, caching strategies proposed in the literature generally aim at optimizing cost-blind metrics (e.g., hit-ratio, number of hops or delay of the request path, link utilization, etc.) that are however uninformative of the cost incurred by the ISP to retrieve content for its users. Ignoring these aspects may incur in additional costs for ISPs, that instead could be avoided under a cost-aware caching design. Distributed implementation of strategies achieving solutions structurally close to MIN-COST is part of our ongoing work.
\label{sec:conclusion}

\section*{Acknowledgment}
\noindent  
This work was partially funded by the DIGITEO project Odessa-CCN, the ANR project Green-Dyspan, and partly carried out at LINCS.


\bibliographystyle{IEEEtran}
\bibliography{IEEEabrv,library,dario}


\end{document}